\documentstyle[aps,prb,twocolumn]{revtex}
\draft
\begin{document}
\title{Interband absorption and luminescence in small 
 quantum dots under strong magnetic fields}
\author{Augusto Gonzalez\cite{augusto}}
\address{Centro de Matem\'aticas y F\'\i sica
 Te\'orica, Calle E No. 309, Ciudad Habana, Cuba and\\
 Departamento de F\'{i}sica, Universidad de Antioquia, AA 1226, 
 Medellin, Colombia\\ and}
\author{Eduardo Men\'{e}ndez-Proupin\cite{eduardo}}
\address{Instituto de Materiales y Reactivos, 
 Universidad de La Habana San L\'{a}zaro y L, Vedado 10400, 
 Ciudad Habana, Cuba}
\date{\today}
\maketitle

\begin{abstract}
Interband absorption and luminescence of quasi-two-dimensional, circularly
symmetric, $N_{e}$-electron quantum dots are studied at high magnetic
fields, $8\leq B\leq 60$ T, and low temperatures, $T\ll 2$ K. In the $%
N_{e}=0 $ and 1 dots, the initial and final states of such processes are
fixed, and thus the dependence on $B$ of peak intensities is monotonic. For
larger systems, ground state rearrangements with varying magnetic field 
lead to substantial modifications of the absorption and luminescence 
spectra. Collective effects are seen in the $N_{e}=2$ and 3 dots at 
``filling fractions'' 1/2, 1/3 and 1/5.
\end{abstract}

\pacs{PACS numbers: 71.35.Ee, 78.66.-w\\
Keywords: quantum dots, electron-hole systems, high magnetic fields}

\section{Introduction}

The quasi-two-dimensional electron gas in a high magnetic field is a
strongly correlated system exhibiting very complicated dynamics. At special
values of the filling factor, the essential features of the ground state are
captured by the Laughlin wave function \cite{Laughlin}, or its composite
fermion generalization\cite{CF}. The low-lying excitations can be described
in the single-mode approximation of Girvin {\it et al},
\cite{GMP,Chakraborty} or in the composite fermion picture \cite{CF,CF2}.

Many experiments have been designed and carried out in order to test the
excitation spectrum of this highly correlated system. Inelastic (Raman)
light scattering experiments have tested basically the excitation gap at
wavevector ${\bf k}=0$.~\cite{Raman} Spin-flipped states and the 
magnetoroton
minimum at $k\approx 1/l_{B}$ ($l_{B}$ is the magnetic length) have also
been observed , although they should be activated by impurities or other
mechanism to produce a trace in the Raman spectra. Evidence of the
magnetoroton minimum comes also from the absorption of ballistic acoustic
phonons\cite{fonones}.

On the other hand, experiments on photoluminescence (PL) related to
interband electronic transitions around filling factor $\nu =1$ have tested
the excited states with an additional electron-hole (e-h) pair\cite{nu1}.
Recently, the observations have been extended to lower filling factors by
increasing the magnetic field up to 60 T.~\cite{Xminus1,Xminus} The related
theory is not in complete agreement with the experiment. In the infinite
magnetic field limit, it was predicted that only the exciton ($X^{0}$) and
the negatively charged triplet exciton ($X_{t}^{-}$) are bound \cite{PYM96}.
The latter is expected to be dark in luminescence \cite{PYM96} as a result
of a hidden symmetry related to magnetic translations \cite{LL81}. In the
experiments, however, very distinct singlet and triplet peaks ($X_{s}^{-}$
and $X_{t}^{-}$) are observed. A realistic calculation of ground state
energies was presented in Ref. \onlinecite{WS97}, where Landau level and
quantum well \ (qwell) sub-band mixing were taken into account. The $%
X_{t}^{-}$ peak position was reproduced, but in theory this state is dark.
The problem was recently revisited by Wojs {\it et al}\cite{Wojs}, who
showed that in a narrow (10 nm wide) well a second bright $X_{t}^{-}$ state
becomes bound, thus interpreting the observed luminescence as coming from
the bright state. We shall notice that both Refs. \onlinecite{WS97} and 
\onlinecite{Wojs} deal with isolated three-particle systems, and thus are
not able to describe the filling factor dependence of observed magnitudes
for $\nu \geq 1/5$.

In the present paper, we study small quantum dots (qdots) under conditions
similar to the experiments reported in Refs. \onlinecite{Xminus1,Xminus}, 
{\it i.e.} quasi-two-dimensional motion, magnetic fields in the interval 
$8$ T$\leq B\leq 60$ T, and temperatures well below 2 K. The laser 
excitation power is
assumed to be low (a few mW/cm$^{2}$), thus the dot works under a linear
regime. The lateral confinement is modelled by a harmonic potential.
Energy levels, charge densities and dipole matrix elements for absorption
and luminescence are computed by exact diagonalization in the first Landau
level (1LL) approximation.

Absorption or PL experiments on electron-hole qdots under very high 
magnetic fields are lacking. To the best of our knowledge, there is only 
one experiment \cite{CRL99}, in which the luminescence at higher (4 K)
temperature and $B\leq 45$ T is measured in order to estimate the e-h
correlation energy.

Breaking of the magnetic translation symmetry by a lateral confinement in a
qdot makes the lowest $X^-$ triplet state bright. Highly nontrivial PL
and absorption spectra arise even in the 1LL approximation. These spectra
contain information about the energy levels and particle correlations in 
the system. Let us stress that a calculation of $X^0$ and $X^-$ energy 
levels of in a qdot, which includes LL mixing, is available \cite{WH95}. The
absorption coefficient is also reported in that paper. The differences with
our work are the following. First, we consider both absorption and PL.
Second, we trace the changes in the ground-state (g.s.) wave function and
charge rearrangements as the magnetic field is varied. Finally, we consider
larger qdots with $X^{2-}$ and  $X^{3-}$ complexes (unbound in a qwell). It 
will be seen below that indications of collective effects are evident even 
in these relatively small systems.

The plan of the paper is as follows. The model and certain general
statements are explained in section II. The next section presents results
for particular systems. We start with the exciton and end up with the 
$X^{3-} $ complex. Finally, a few concluding remarks are given.

\section{The model}

We consider the two-dimensional motion of $N_e$ electrons and $N_h$ holes 
in an external parabolic potential and a perpendicular magnetic field 
(along the $z$ axis). In particular, we will study the $N_h=1$ and 2 
systems, which
are the ones participating in interband absorption and recombination
processes. The unit of length is $\sqrt{2}$ times the magnetic length. In
the 1LL approximation, the Hamiltonian is written as

\begin{eqnarray}
H(N_e,N_h)&=&\left (\frac{\hbar\omega_c^e}{2}+E_z^e\right )N_e+ \left (\frac{%
\hbar\omega_c^h}{2}+E_z^h+E_{gap}\right )N_h  \nonumber \\
&+& E_{Zeeman}^e+E_{Zeeman}^h+V_{conf}+V_{coul}.  \label{hamiltoniano}
\end{eqnarray}

\noindent 
The Hamiltonian (\ref{hamiltoniano}) is intended to model a GaAs qdot
with a thickness of 20 nm in the $z$-direction. The meaning of the 
different terms entering $H$ is evident.
The specific qdot characteristics are reflected in the confinement energies
along the $z$-direction, $E_{z}$, the in-plane confinement potential, 
$V_{conf}=\sum_{i}v_{conf}(r_{i})$, and the $z$-averaged Coulomb 
interactions, $V_{coul}=\sum_{i,j}v_{coul}(r_{ij})$. We will use the 
expression

\begin{equation}
v_{coul}(r)=\pm 3.316~\beta~\sqrt{B}~\frac{1}{r}~[{\rm meV}],
\end{equation}

\noindent 
for the pair Coulomb interactions ($B$ in Teslas), and

\begin{equation}
v_{conf}(r)=\frac{3.316}{B}~\beta~K~r^2~[{\rm meV}],
\end{equation}

\noindent 
for the one-particle confinement potential. Even these simple expressions
lead to very interesting physical results. Notice that $1/\sqrt{2}$ times
the characteristic 
Coulomb energy, $e^2 /(\kappa l_B)$, equals 3.316 $\sqrt{B}$ in our units.
The constant $\beta =0.6$ is used
to simulate the $z$-averaging of the Coulomb interactions in the 20 nm -
wide qdot \cite{Chakraborty,MA84}. We fixed it by requiring the binding
energy of the unconfined ($v_{conf}$ set to zero) $X_{t}^{-}$ relative to
the $X^{0}$ to be 0.6 meV at $B=17$ T. This is a representative value\cite
{Xminus}. On the other hand, the dimensionless constant $K$ will be fixed 
to 7.0 in order to obtain a ``filling factor'' around 1/3 for 
$B\approx 30$ T, also a common situation met in the experiments 
\cite{Xminus}.

The only nontrivial terms entering (\ref{hamiltoniano}) are $V_{conf}$ and 
$V_{coul}$. They should be diagonalized in a basis of Slater 1LL functions.
The energies coming from the diagonalization processes will be denoted 
$\epsilon$, and the wave functions will be used to compute physical
observables. Note that, in the 1LL, the electron (hole) angular momentum is
a non-positive (non-negative) number. Thus, the total angular momentum is
written $M=M_e+M_h=-|M_e|+M_h$.

In a GaAs electron system, the validity of the 1LL approximation can be
stated by comparing the excitation energy to the 2LL, $\hbar \omega
_{c}^{e}=1.728~B$ meV, with the Coulomb energy, $3.316~\beta ~\sqrt{B}$
meV. Thus, for $B\gg 1$ T the 1LL approximation works. Spin excitations
are lower in energy, $\Delta E_{Zeeman}\approx 0.025~B$ meV. However, at
temperatures below 2 K and for $B>8$ T, they can not be thermally excited.
It means that in both absorption and luminescence the transition starts 
from the lowest optically active state. When holes are created, the 1LL
approximation becomes valid at higher fields. If we take for the heavy hole
mass in the $xy$ plane the value $\mu_{h}=0.11~m_{0}$\cite{mh}, then $\hbar
\omega _{c}^{h}\approx 1B$ meV. The 1LL approximation works for $B\gg 4$ T.
Below, we present results obtained in the 1LL approximation for $8$ T$\leq
B\leq 60$ T.

On the other hand, the expression (\ref{hamiltoniano}) assumes that the
particles are sitting on the first qwell sub-band. As it was stressed in
Ref. \onlinecite{WS97}, this may be a rough approximation. For a 20 nm
qwell, the second electronic sub-band is around 30 meV higher, but the
second hole sub-band is only 6 meV higher (a heavy hole mass 
$\mu_{h}^{z}\approx 0.38~m_{0}$ is assumed). Our first sub-band approximation
is qualitatively valid in the present situation, and will improve for
narrower wells.

\subsection{Interband absorption and luminescence (general grounds)}

Interband absorption and luminescence will be studied at temperatures 
$T\ll 2$ K, {\it i.e.} typically lower than spin excitation gaps. Thus, the 
processes proceed from a unique initial state, which is the g.s. of the 
polarized $(N_{e},0)$ system in absorption, and the lowest optically 
active state of the $(N_{e}+1,1)$ system in emission. In general, these 
processes take place
in different angular momentum channels. For absorption, the incident light
is supposed to be circularly polarized and propagating along the 
$z$-direction. Also circularly polarized light is supposed to be measured 
from the qdot luminescence.

A simple two-band model, with bands split by the Zeeman energy, will be
used. The conduction-band ($m_{s}=\pm 1/2$) mass is $\mu_{e}=0.067~m_{0}$, 
and the heavy hole band, $m_{j}=\pm 3/2$, shows anisotropic effective 
masses, $\mu_{h}=\mu_{h}^{xy}=0.11~m_{0}$, $\mu_{h}^{z}=0.38~m_{0}$. 
LL mixing 
in the $m_{j}=3/2$ branch \cite{RLZ59} will be neglected. $m_{j}=-3/2$ will 
be called the spin-up hole branch, and $m_{j}=3/2$ -- the spin-down branch. 
For propagation along the $z$ axis, the allowed transitions are 
$m_{j}=-3/2\rightarrow m_{s}=-1/2$ for right-handed circular polarization
(RHCP), and $m_{j}=3/2\rightarrow m_{s}=1/2$ for left-handed circular
polarization (LHCP) \cite{RLZ59,SPRS95}.

The dipole approximation is used for the interaction Hamiltonian, {\it i.e.} 
$-{\cal E} \cdot {\bf D}$. In the 1LL, the interband dipole operator 
takes the form

\begin{equation}
{\bf D}=\frac{e{\bf p}_{cv}}{m_{0}\omega }\sum\limits_{l\geq 0}\left(
e_{-l,\downarrow }^{\dagger }h_{l,\uparrow }^{\dagger }+e_{-l,\uparrow
}^{\dagger }h_{l,\downarrow }^{\dagger }\right) +H.C.,  \label{hint}
\end{equation}

\noindent 
where ${\bf p}_{cv}$ is the GaAs interband constant. The reason
for not including the light hole in (\ref{hint}) is twofold. First, $E_{z}$
is around 6 meV higher ($\mu_{lh}^{z}\approx 0.09~m_{0}$), thus its
absorption or luminescence lines are shifted. Second, the constant 
${\bf p}_{cv}^{2}$ is three times smaller for light holes. Notice that the
interaction Hamiltonian preserves total angular momentum.

In our $(N_e,N_h)$ systems with $N_h=0,~1$, the states may be classified
according to the symmetry of the electronic subsystem. For example, the 
$N_e=2$ system may be in a spatially antisymmetric (triplet) state, or in a
spatially symmetric (singlet) state. We will present calculations only for
spatially antisymmetric states. They are the only ones appearing in LHCP,
and the ones associated to the most intense lines in RHCP \cite{WH95}. The
wave functions may be written as $\psi=\phi_{coord}^{antisymm}
\chi_{spin}^{symm}$, or in a second quantization formalism,

\begin{equation}
\left| \psi (N_{e},0)\right\rangle =\sum C_{l_{1}l_{2}\dots
l_{N_{e}}}e_{-l_{1},\uparrow }^{\dagger }e_{-l_{2},\uparrow }^{\dagger
}\dots e_{-l_{N_{e}},\uparrow }^{\dagger }|0\rangle ,
\end{equation}

\begin{eqnarray}
&&\left| \psi _{LHCP}(N_{e}+1,1)\right\rangle =  \nonumber \\
&&\sum C_{l_{1}l_{2}\dots l_{N_{e}+1},l_{h}}e_{-l_{1},\uparrow }^{\dagger
}e_{-l_{2},\uparrow }^{\dagger }\dots e_{-l_{N_{e}+1},\uparrow }^{\dagger
}h_{l_{h},\downarrow }^{\dagger }|0\rangle ,
\end{eqnarray}

\begin{eqnarray}
&&\left| \psi _{RHCP}(N_{e}+1,1)\right\rangle  =\frac{1}{\sqrt{N_{e}}}\sum
C_{l_{1}l_{2}\dots l_{N_{e}+1},l_{h}}  \nonumber \\
&&\times \left( e_{-l_{1},\downarrow }^{\dagger }e_{-l_{2},\uparrow
}^{\dagger }\dots e_{-l_{N_{e}+1},\uparrow }^{\dagger }\right.\nonumber \\
&&+e_{-l_{1},\uparrow }^{\dagger }e_{-l_{2},\downarrow }^{\dagger }\dots
e_{-l_{N_{e}+1},\uparrow }^{\dagger }  \nonumber \\
&&+\dots +\left. e_{-l_{1},\uparrow }^{\dagger }e_{-l_{2},\uparrow
}^{\dagger }\dots e_{-l_{N_{e}+1},\downarrow }^{\dagger }\right)
h_{l_{h},\uparrow }^{\dagger }|0\rangle .
\end{eqnarray}

\noindent 
$\psi _{LHCP}$ corresponds to a spin-polarized electronic
subsystem, and $\psi _{RHCP}$ to a not completely polarized state. In the
pure electron system, the sum runs over angular momentum states obeying 
$0\leq l_{1}<l_{2}<\dots <\l _{N_{e}}$ and fixed $M=-l_{1}-l_{2}-\dots
-l_{N_{e}}$. In the one-hole system, the total angular momentum 
$M=-l_{1}-l_{2}-\dots -l_{N_{e}+1}+l_{h}$ is fixed.

Diagonalization of $V_{conf}+V_{coul}$ in (\ref{hamiltoniano}) leads to the
determination of eigenenergies and wave functions. Transition energies,
transition probabilities and charge densities of the relevant states are
computed from these results. The transition energies are given by

\begin{eqnarray}
\hbar\omega&=&E_{gap}+E_z^e+E_z^h+\frac{\hbar\omega_c^e}{2}+ 
\frac{\hbar\omega_c^h}{2}  \nonumber \\
&+& E_{Zeeman}^e+E_{Zeeman}^h+ \epsilon (N_e+1,1)-\epsilon (N_e,0),
\label{homega}
\end{eqnarray}

\noindent 
where $\epsilon $ are the energies coming from $V_{conf}+V_{coul}$. 
We took the values $E_{gap}=1510$, $E_{z}^{e}=11$, $E_{z}^{h}=2$, 
$\hbar \omega _{c}^{e}/2=0.864~B$, $\hbar \omega _{c}^{h}/2=0.526~B$, 
$E_{Zeeman}^{e}=-0.025~m_{s}^{e}~B$, $E_{Zeeman}^{h}=-0.016~m_{s}^{h}~B$, 
for the quantities entering (\ref{homega}), where energies are given in meV 
and $B$ in Teslas. Our treatment of Zeeman energies of both electrons and 
holes is very simple. We used the value $g_{e}=-0.44$ for the electron 
Land{\'e} factor and extracted the hole energy from the observed splitting of 
$X^{0}$ luminescence lines in RHCP and LHCP \cite{Xminus}. The hole spin 
projection is conventionally written as $m_{s}^{h}=\pm 1/2$. Actually, the 
Zeeman energy shows a nonlinear dependence on $B$. \cite{SPW97} Notice, 
however, that $E_{gap}$, $\hbar \omega _{c}$ and $E_{Zeeman}$ are important 
in determining the absolute position of a given absorption or PL line, but 
not its relative position with respect to $X^{0}$ in the same polarization.

The absorption coefficient of a dot is given by

\begin{equation}
\alpha (\omega )=\frac{4\pi ^{2}\omega }{\hbar cV}\sum_{f}|\langle f|
{\bf e}\cdot {\bf D}|i\rangle |^{2}\delta (\omega -\omega _{fi}),
\end{equation}

\noindent where $|i\rangle $ is the g.s. of the $(N_{e},0)$ system, $f$ are
the states of the $(N_{e}+1,1)$ system in the same angular momentum tower
and $\hbar \omega _{fi}$ is their energy difference computed from (\ref
{homega}). ${\bf e}$ is the light polarization vector, $c$ -- the
light velocity, and $V$ is the volume of absorption. We have used a 
phenomenological width, $\Gamma =0.8$ meV, to replace the delta function 
by a Lorentzian

\begin{equation}
\delta (x) \to \frac{\Gamma/\pi}{\Gamma^2+x^2}.
\end{equation}

In luminescence, we compute the matrix elements $|\langle f|{\bf e}\cdot 
{\bf D}|i\rangle |^{2}$, assuming that $|i\rangle $ is the lowest state of
the $N_{h}=1$ system.

\section{Results}

We present results in the following interval of magnetic field values, 
8 T $\leq B\leq 60$ T. Computations are carried out for spin polarized
electronic systems, with total spin $M_{s}^{e}=N_{e}/2$, which contribute 
to the LHCP spectra. The energies of the incompletely polarized states with 
$M_{s}^{e}=N_{e}/2-1$, entering the RHCP spectra, are obtained
by adding the corresponding Zeeman shifts.

\subsection{Binding energies of excitonic complexes}

We draw in Fig.~\ref{fig1} the g.s. energies, $\epsilon$, coming from the
diagonalization of $V_{conf}+V_{coul}$ in (\ref{hamiltoniano}) as a 
function of the applied magnetic field. The polarized systems 
$(N_{e}+1,N_{h})$= (1,1),
(2,1), (3,1) and (4,1) are shown. The common notation for the excitonic
systems (1,1) and (2,1) are $X^{0}$ and $X^{-}$, so that the charged 
complex (4,1) may be denoted $X^{3-}$. Note that the slopes of the (2,1), 
(3,1) and
(4,1) curves are very similar. It means that the relative binding energies
vary smoothly with $B$, and that the magnetic moments of these states take
almost the same values. For example, $X^{3-}$ is 14.77 meV above $X^{-}$ at 
$B=30$ T, and 14.29 meV above $X^{-}$ at $B=50$ T.

The total angular momenta in the g.s. is a constant, independent of $B$, in
the smallest systems. It is $M_{gs}=0$ in the exciton, and $M_{gs}=-1$ in
the triplet $X^-$ at any $B$. The larger systems, however, undergo abrupt
rearrangements at particular $B$ values. The interplay between g.s.
rearrangements in the $(N_e,0)$ and $(N_e+1,1)$ systems as $B$ is varied has
direct consequences on absorption and luminescence, as will be seen below.

Note that, unlike pure electron systems, when holes are present the Hilbert
space in a given $M=-|M_{e}|+M_{h}$ sector is not finite. We enlarged the
included subspace until convergence is reached. For example, in the (4,1)
system at $B=40$ T, 2374 many-particle states ({\it i.e.} all of the states 
in $15\leq |M_{e}|\leq 35$) are enough to reach convergence for the lowest
energy eigenvalue in the $M=-15$ tower.

The low-lying energy levels of $X^{3-}$ at $B=35$ T are shown in Fig. 
\ref{fig2} as an example. Energy distances between the lowest adjacent 
levels are around 0.5 meV, the same as in the three-electron system at 
this value of the magnetic field.

\subsection{Interband absorption}

As previously stated, temperatures are low enough for absorption to proceed
from the g.s. of the $N_{e}$-electron system. It means that spin flips
should not be thermally induced, {\it i.e.} $T\ll 2$ K for $B>8$ T.

We show in Fig.~\ref{fig3} the absorption coefficient for the $N_{e}=0$ qdot
at $B=40$ T. The process under consideration, $(0,0)\rightarrow (1,1)$,
goes through the $M=0$ channel. The main properties of the curve drawn in
Fig.~\ref{fig3}, {\it i.e.} dominance of the exciton g.s. and monotony, are
visible also at any other value of the magnetic field. The main effect of 
$B$ is to reinforce the dominance of the first line. The threshold for
absorption is determined by the exciton g.s. energy, and the maximum dipole
squared behaves like $B^{0.78}$.

The absorption coefficient of the negatively charged dot, $N_{e}=1$, is
shown in Fig.~\ref{fig4}. The $(1,0)\rightarrow (2,1)$ process takes place
in the $M=0$ sector. At $B=8$ T, a structure of isospaced
bands is seen in the spectrum at higher energies. Most of these lines are
suppressed already at $B=40$ T. The threshold for absorption and maximum
strength transition are determined by the lowest $X^{-}$ state in the $M=0$
tower. As a function of $B$, we get $D^{2}\sim B^{0.79}$ at the maximum.

The absorption thresholds for the smallest systems, $N_{e}=0$ and 1, are
smooth functions of $B$, signalling that the states entering the transition $%
(N_{e},0)\rightarrow (N_{e}+1,1)$ do not change qualitatively as $B$ is
raised. For larger systems, however, there is an abrupt decrease in the
threshold for fields around 10 T (``filling factor'' near one), and small
steps at higher fields . The steps are originated by the different rates of
change of $M_{gs}$ in the $(N_{e},0)$ and $(N_{e}+1,1)$ systems (see Table
\ref{tab1}). Let us
consider, for example, the $(3,0)\rightarrow (4,1)$ process. For $B\leq 10$
T, the process goes from the g.s. of (3,0) to the excited states of $X^{3-}$
with $M=-3$. For $B>10$ T, the g.s. of (3,0) moves to $M=-6$, a sector
which contains the g.s. of (4,1). Thus, the threshold is lowered. Every 
time one of the systems rearranges, there is a step like change in the 
absorption threshold. The actual (experimental) profile is expected to be 
smoothed because of temperature effects.

Of course, not only the threshold changes, but the whole spectrum is
restructured. We show in Fig.~\ref{fig5} the absorption in the $N_{e}=2$ 
dot ($X^{2-}$ formation) at $B=8$ T and 50 T. At $B=8$ T, the spectrum
is similar to the $X^{-}$ spectrum. The added electron is placed in an outer
orbit because the inner orbitals are filled. For higher fields, there is
place for the new electron in the core region, but the minimization of
energy causes a global restructuration of the charge density in the dot, as
will be seen below. The added pair losses its identity.
Notice that for $B>10$ T there are two very distinct lines in the spectrum.
One is the threshold (the transition to the lowest state of (3,1)), and the
second is the maximum, which is 7-4 meV above the threshold. 

The dipole squared at maxima as a function of $B$ are drawn in Fig.
~\ref{fig6}. Besides
lowest state rearrangements, there are manifestations of collective effects
even in these small systems. A decrease of absorption in the $N_{e}=2$ and 3
systems at ``filling factors'' $\nu \approx 1/2$, 1/3 and 1/5 is evident
from Fig.~\ref{fig6}.

\subsection{Magnetoluminescence}

The second part of  Fig.~\ref{fig5} shows the square of the dipole matrix 
elements
corresponding to the luminescence of the $N_{e}=2$ dot at $B=40$ T. Only
transitions starting from the g.s. of (3,1) are considered. Notice that the
lowest state of (2,0) gives the strongest line, approximately 50 times
higher than the next one. This is the common situation in our luminescence
calculations for any of the systems under study. The strongest line
corresponds to the transition from the g.s. of $(N_{e}+1,1)$ to the lowest
state of $(N_{e},0)$ in the same angular momentum tower. The higher states
of $(N_{e},0)$ give negligible contributions.

Luminescence in the $N_{e}=0$, and 1 dots is monotonic with $B$ because the
initial and final states participating in it are fixed. Exciton luminescence
proceeds in the $M=0$ channel, and $X^{-}$ luminescence in the 
$M=-1$ sector. In the latter case, the absorption and luminescence channels
are different. With increasing $B$, the $X^{0}$ peak intensity increases, 
as in absorption, but the $X^{-}$ intensity
decreases. We obtained $D^{2}\sim $ Exp$(-0.018~B)$ at the maximum. 

For larger systems, the luminescence shows non monotonic behaviour because 
of lowest state rearrangements and collective effects, as in absorption. 
As a rule, the channels for absorption and PL are different in these 
systems. The luminescence maxima as a function of $B$ are drawn in 
Fig.~\ref{fig7}.

\subsection{Charge densities}

Electron and hole charge densities inside the dot for the relevant states
participating in absorption and luminescence are presented in this section.
For electrons, we found more convenient to draw the difference 
$\rho_e^{\prime}=\rho_e(N_e+1,1)-\rho_e(N_e,0)$, which gives the density
``added'' to the dot.

Figure \ref{fig8} shows the final-state densities in the absorption 
situations discussed in Fig.~\ref{fig5}. For the $N_e=2$ dot at $B=8$ T, 
the added electron and hole densities are almost identical. The exciton 
keeps its identity inside the dot. At  $B=50$ T, however, the
added pair causes a redistribution of the charge density of the initial
two-electron state.

On the other hand, as shown above, the relevant states participating in
luminescence transitions are the g.s. of $(N_{e}+1,1)$ and the lowest state
of the $N_{e}$-electron system in the same angular momentum sector. We
show in Fig.~\ref{fig9} the densities of these states in the $N_{e}=2$ dot
at $B=40$ T. These curves are typical. The exciton is annihilated from a 
distribution very similar to the isolated exciton g.s. 
(also shown in the figure for comparison).

\section{Concluding remarks}

We have studied few-electron systems and excitonic complexes (with one 
hole) in qdots under intense magnetic fields and low temperatures.  
In 1- and 2-electron qdots the g.s. angular momentum is 
independent of the magnetic field intensity. However, larger systems 
undergo abrupt rearrangements at particular $B$ values, a fact that is 
reflected in the optical absorption and PL.

We computed the interband optical properties of these systems. 
In absorption, the initial state is the polarized ground state of 
$N_{e}$ electrons (for temperatures $\ll 2$ K), and the final states 
are the states of $N_{e}+1$ electrons and one hole. The
main result of these computations is the non monotonic behaviour of the
absorption maxima 
in the larger ($N_{e}=2$ and 3) systems as the field is
varied (Fig.~\ref{fig6}). This result can be understood as a consequence of
ground state rearrangements and collective effects. We have presented 
typical  charge densities in support of this picture.  
We found a reduction of absorption at ``filling factors'' 1/2, 1/3 and 1/5.

For luminescence events, we have considered the recombination from the 
g.s. of $N_{e}+1$ electrons and a hole. At a given magnetic field intensity,  
the angular momentum of this state may be different from the 
$N_{e}$-electron g.s. angular momentum.
Thus, intrinsic absorption and luminescence may proceed through different
channels. Of particular interest is that, opposite to the qwell case, the
ground state of the 
negatively charged exciton $X_t^{-}$ is bright in luminescence. This is a
consequence of the qdot lateral confinement. Furthermore, for very high $B$
the $X_t^{-}$ state recovers its dark character as compared with the other
complexes here studied. On the other hand, the maximum of the recombination
oscillator strength is a monotonic function of $B$ for qdots with 1 or 2
electrons and a hole, but it is nonmonotonic for qdots with more electrons,
showing collective effects even in these small dots.

Although our calculations for finite systems with a smooth lateral
confinement can not be easily extrapolated to the infinite limit, our
results suggest that many-body effects should be taken into account in the
computation of the $X^{-}$ luminescence in a qwell. Whittaker and Shields 
\cite{WS97}, and Wojs {\it et al} \cite{Wojs} have used a three-particle
model for the $X^{-}$. This model is indeed useful at very high magnetic
fields. At intermediate values of $B$, the magnetoexciton size,
which is $\sim 2~l_{B}\sim 50/\sqrt{B}$ nm, becomes comparable to the 
inter-electronic distance, around 20 nm for a typical carrier density of 
1-2$ \times 10^{11}$ cm$^{-2}$. Many-body effects should take care of the
observed dependence of the PL maximum with the filling factor.

We have not attempted a more sophisticated calculation in these systems
because of the absence of experimental results for qdots in very intense
magnetic fields. Nevertheless, our simple approach (1LL, one qwell sub-band,
parabolic lateral confinement, unrealistic Zeeman energies and $z$-averaged 
Coulomb interactions) captures the essential physics and indicates the 
importance of collective effects even in small qdots.

\acknowledgements
A. G. acknowledges support by the Caribbean Network for Theoretical 
Physics. E. M-P acknowledges the Abdus Salam ICTP, where part
of this work was done. The authors are grateful to C. Trallero-Giner for 
many useful discussions.

\begin{figure}
\caption{G.s. energies of the excitonic $X^{N_e-}$ complexes, denoted also
 $(N_e+1,1)$ in the main text.}
\label{fig1}
\end{figure}

\begin{figure}
\caption{Low-lying energy levels of the polarized $X^{3-}$ complex at 
 $B=35$ T.}
\label{fig2}
\end{figure}

\begin{figure}
\caption{Absorption coefficient of the neutral ($N_e=0$) qdot at $B=40$ T.}
\label{fig3}
\end{figure}

\begin{figure}
\caption{Absorption coefficient of the $N_e=1$ qdot at $B=8$ and 40 T.}
\label{fig4}
\end{figure}

\begin{figure}
\caption{Absorption and PL in the $N_e=2$ dot.}
\label{fig5}
\end{figure}

\begin{figure}
\caption{Squared dipole matrix elements of the strongest absorption lines 
 in the $N_e$-electron qdots vs. $B$.}
\label{fig6}
\end{figure}

\begin{figure}
\caption{Luminescence maxima in the $N_e$-electron qdots 
 vs. $B$.}
\label{fig7}
\end{figure}

\begin{figure}
\caption{Charge densities in final states with maximal absorption. The same
cases as in Fig.~\ref{fig5} are considered.}
\label{fig8}
\end{figure}

\begin{figure}
\caption{Charge densities of the state with maximal oscillator strength in
the luminescence of the $N_e=2$ qdot at $B=40$ T.}
\label{fig9}
\end{figure}

\begin{table}
\caption{Ground-state orbital angular momentum in the $N_e$=2  and 
 3 dots.}
\label{tab1}
\begin{tabular}{|l|l|l|l|l|}
$B [T]$ & $M_{gs}(2,0)$ & $M_{gs}(3,1)$ & $M_{gs}(3,0)$ & $M_{gs}(4,1)$\\
\hline\hline
8   & 1  & 3  & 3  & 6  \\
16  & -  & -  & 6  & 6  \\
20  & 3  & 3  & 6  & 9  \\
25  & -  & -  & 9  & 9  \\
30  & 3  & 3  & 9  & 12 \\
35  & 5  & 3  & 12 & 12 \\
40  & 5  & 5  & 12 & 15 \\
45  & -  & -  & 15 & 15 \\
50  & 7  & 5  & 15 & 18 \\
58  & 7  & 7  & -  & -  \\
\end{tabular}
\end{table}

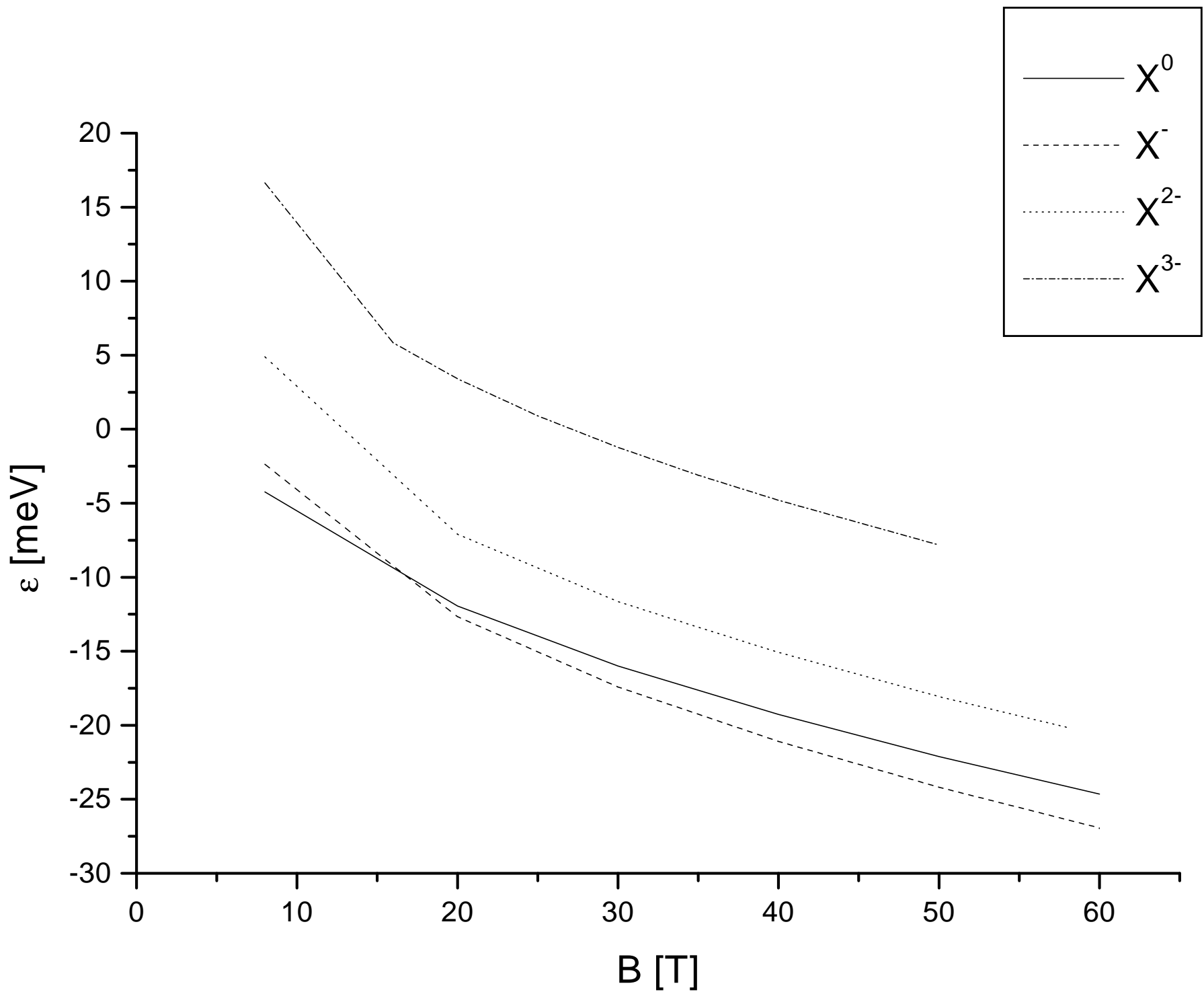
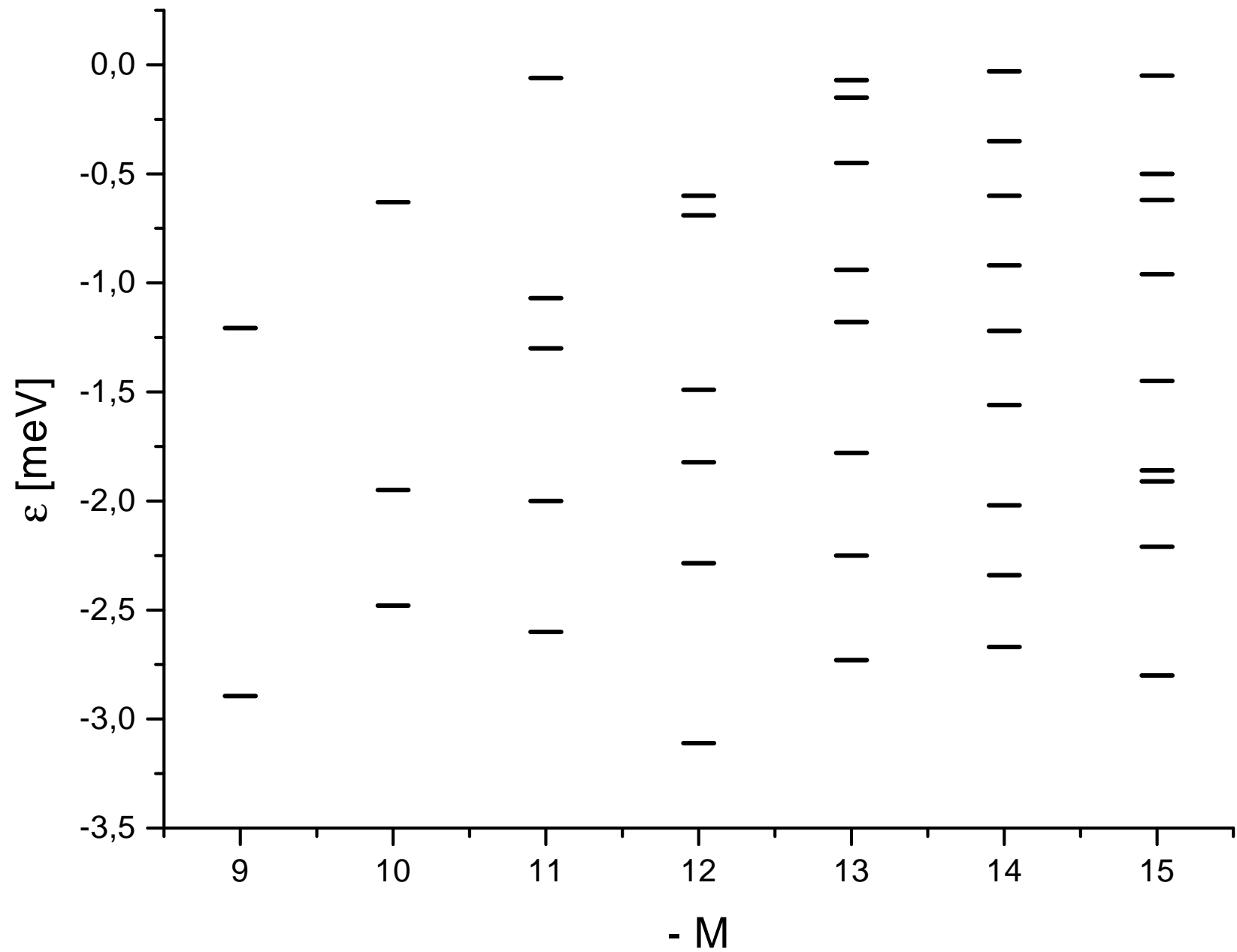
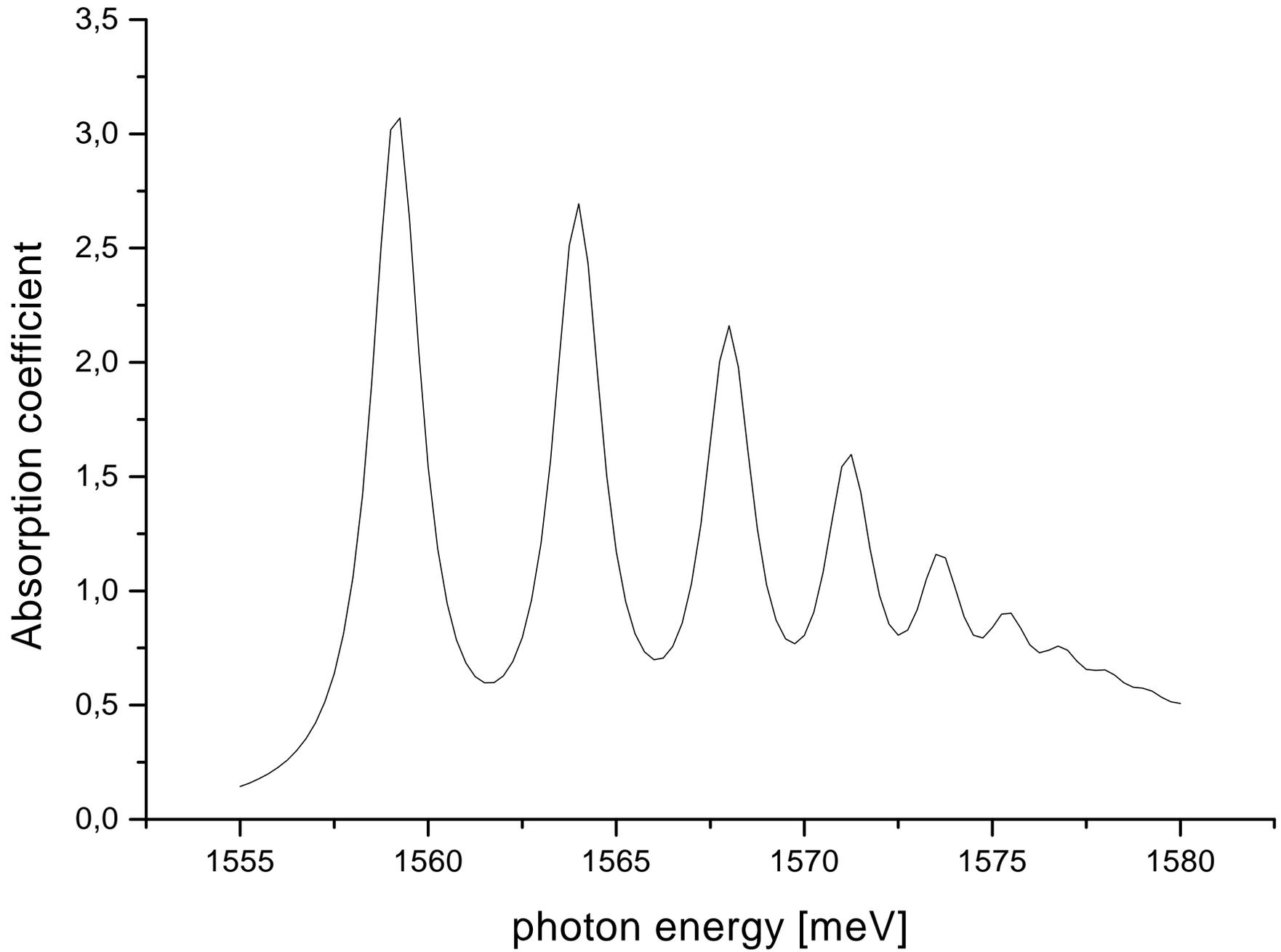
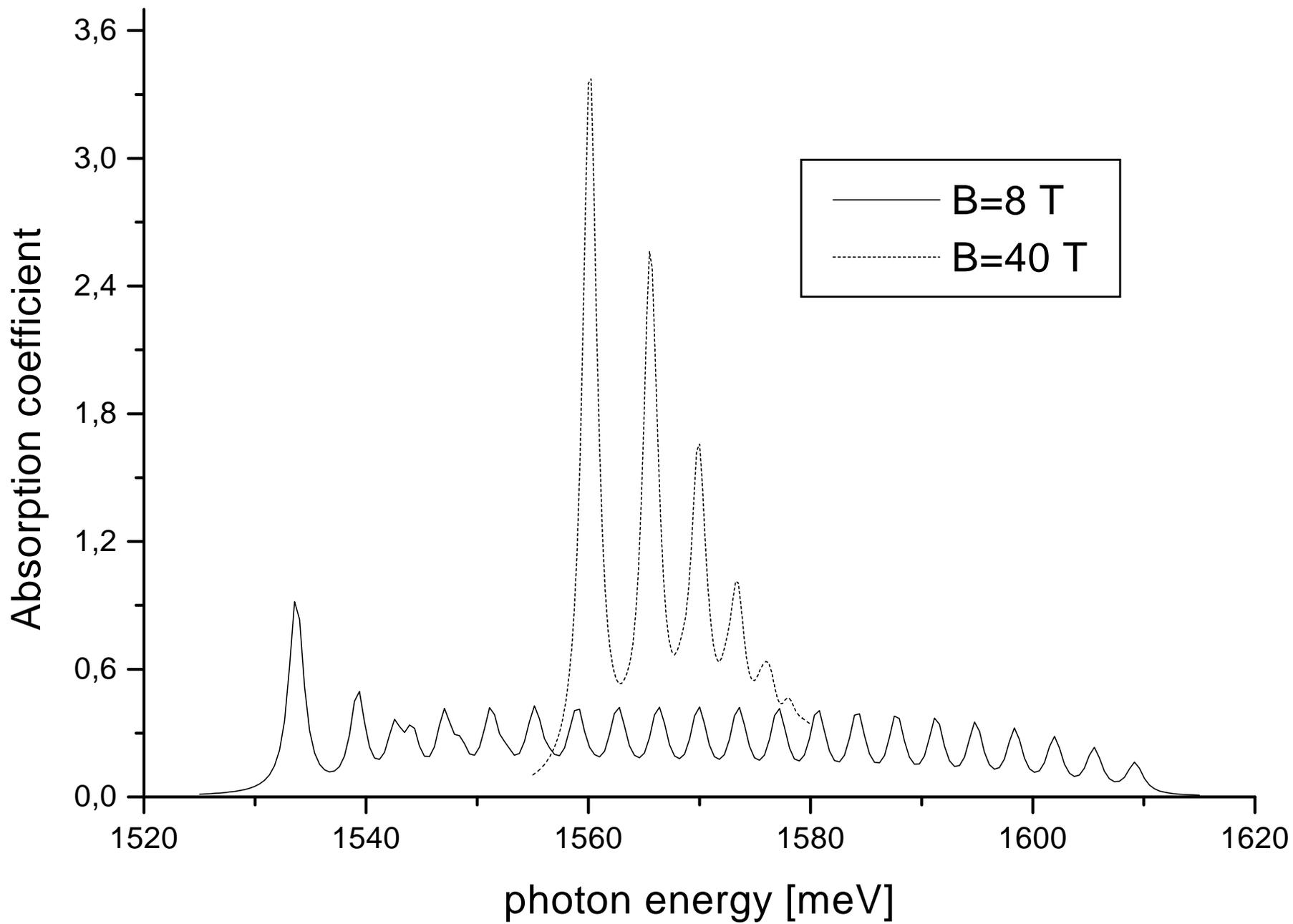
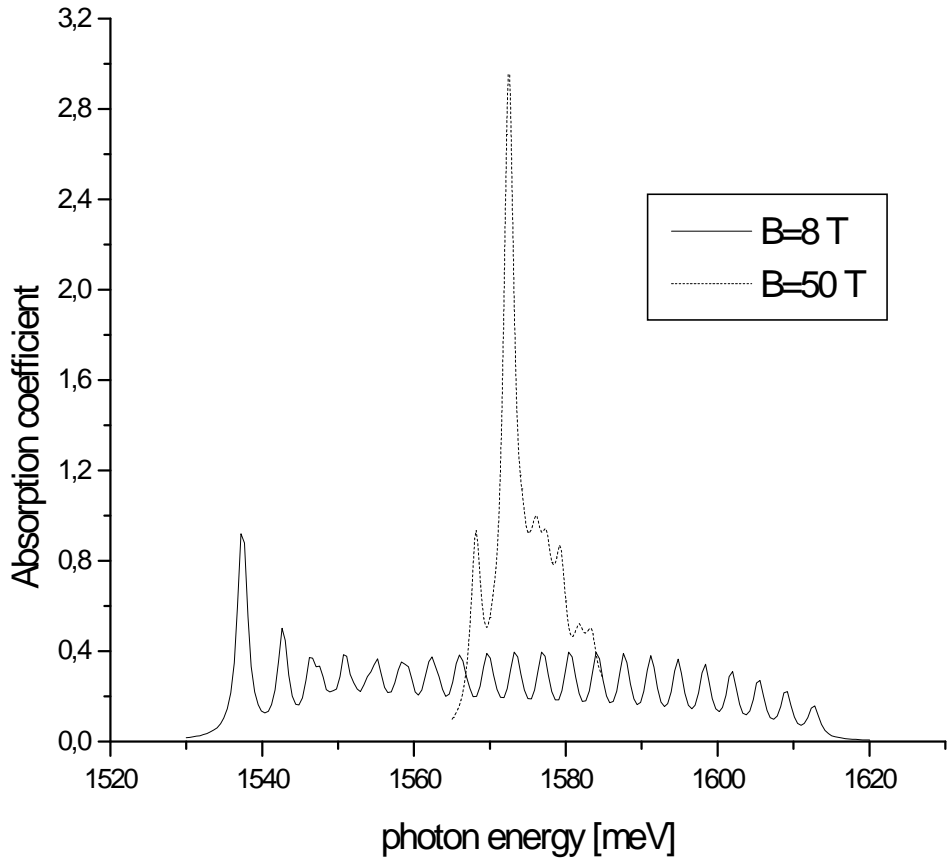
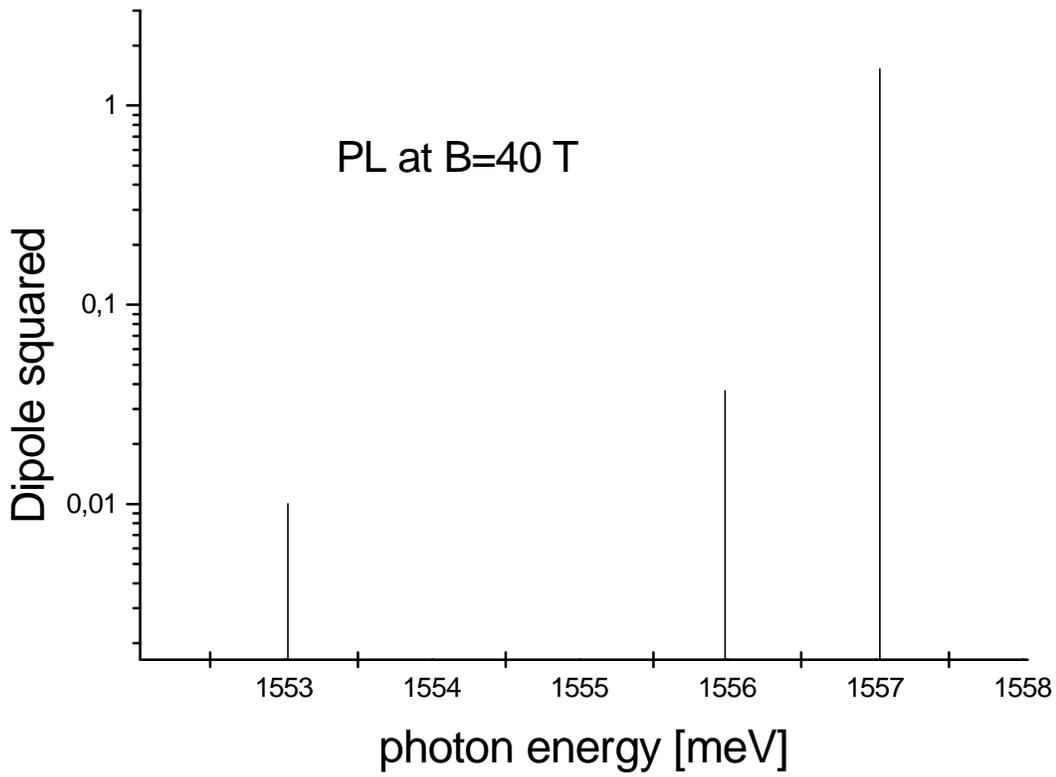
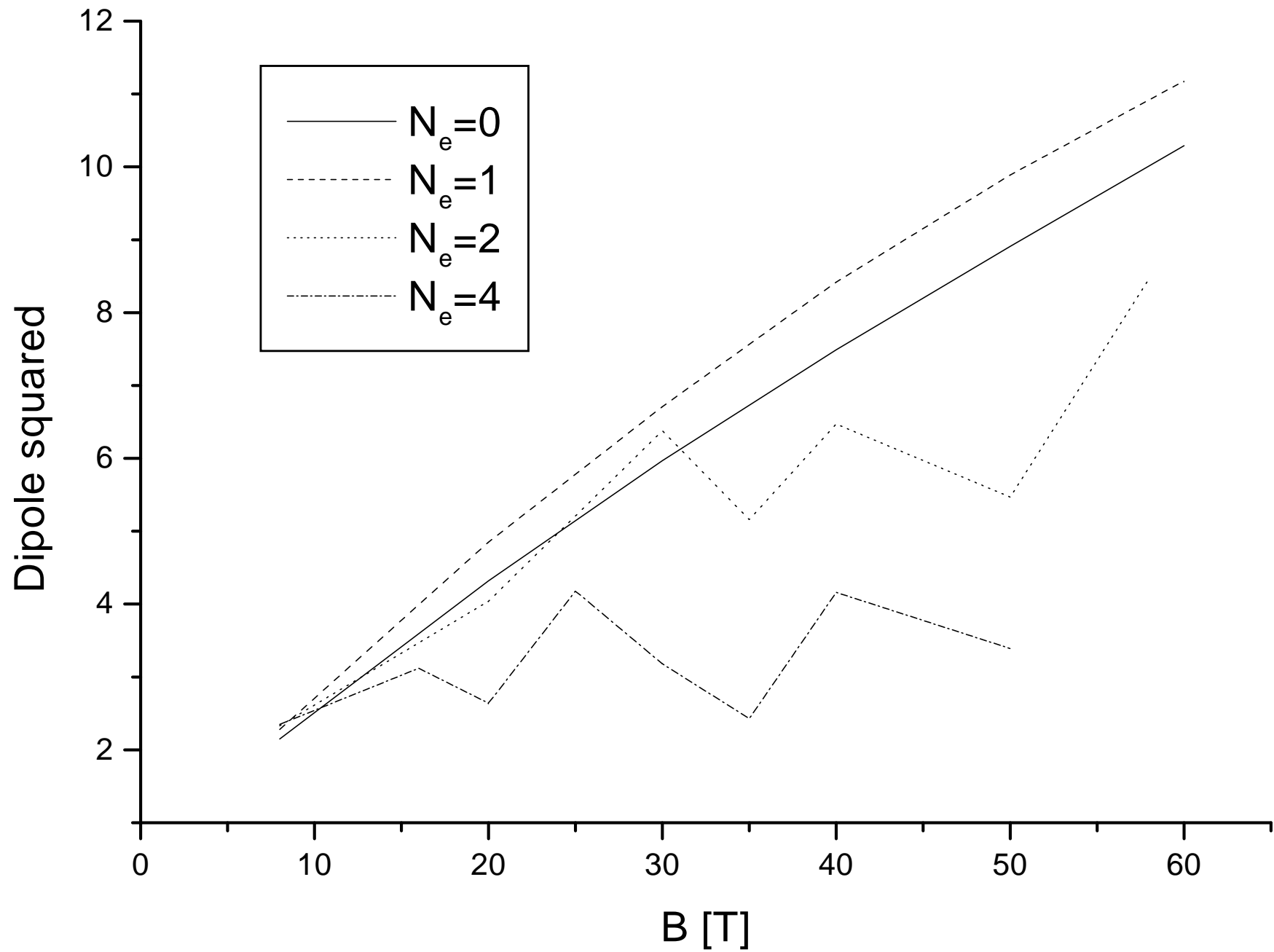
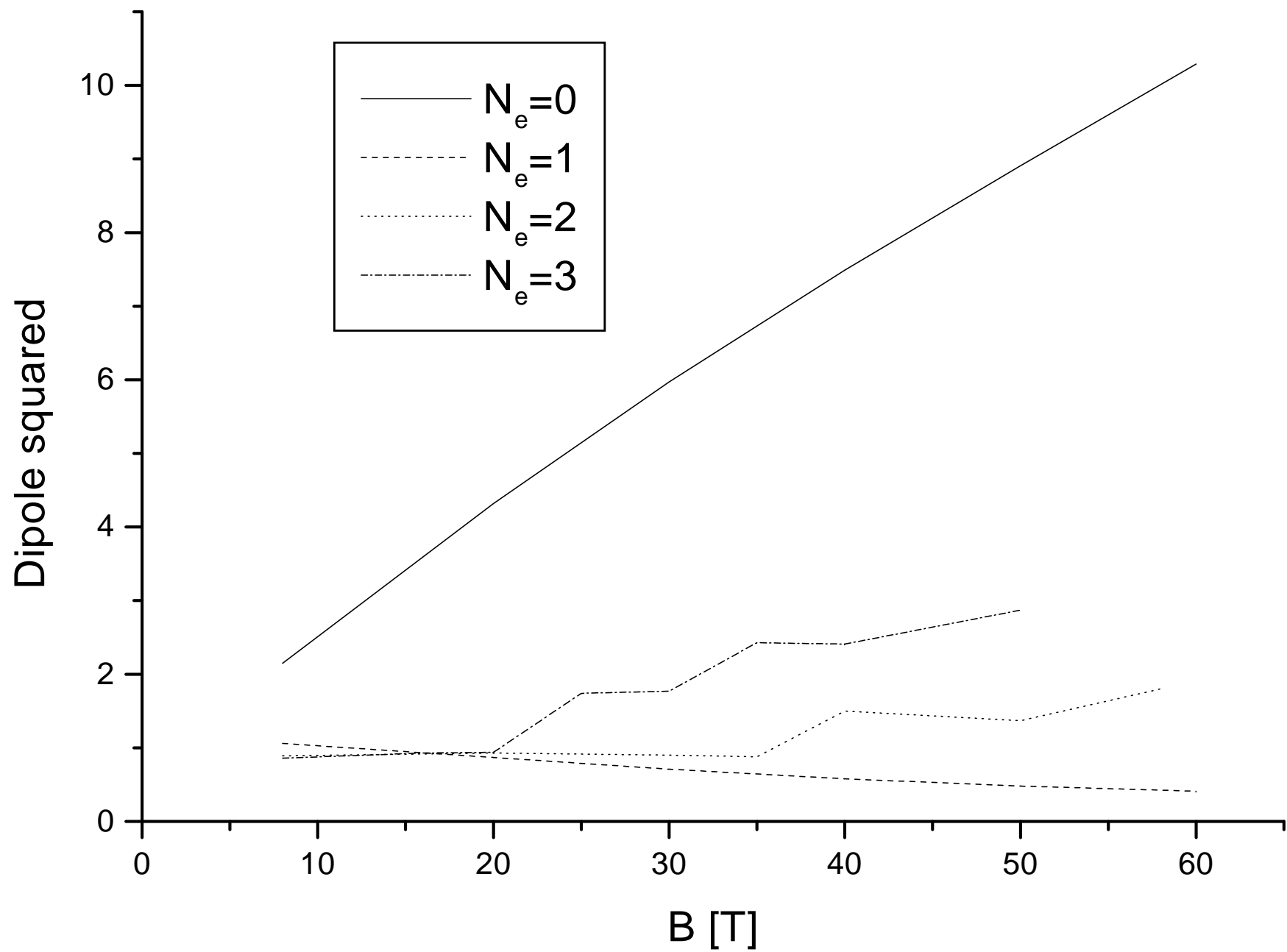
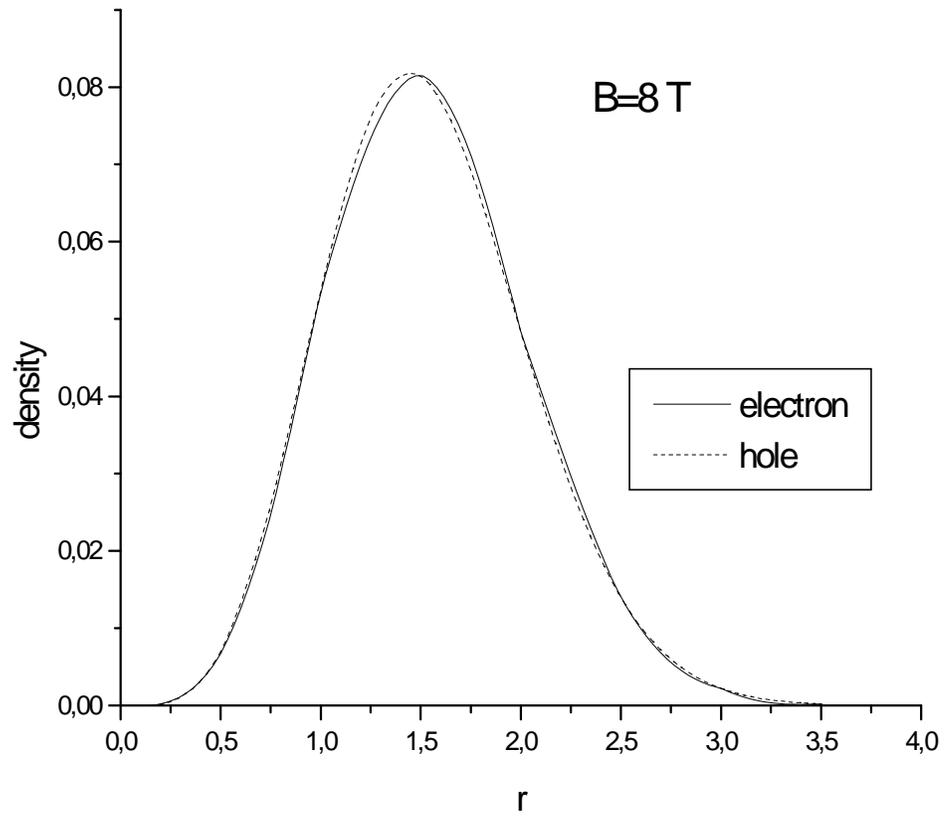
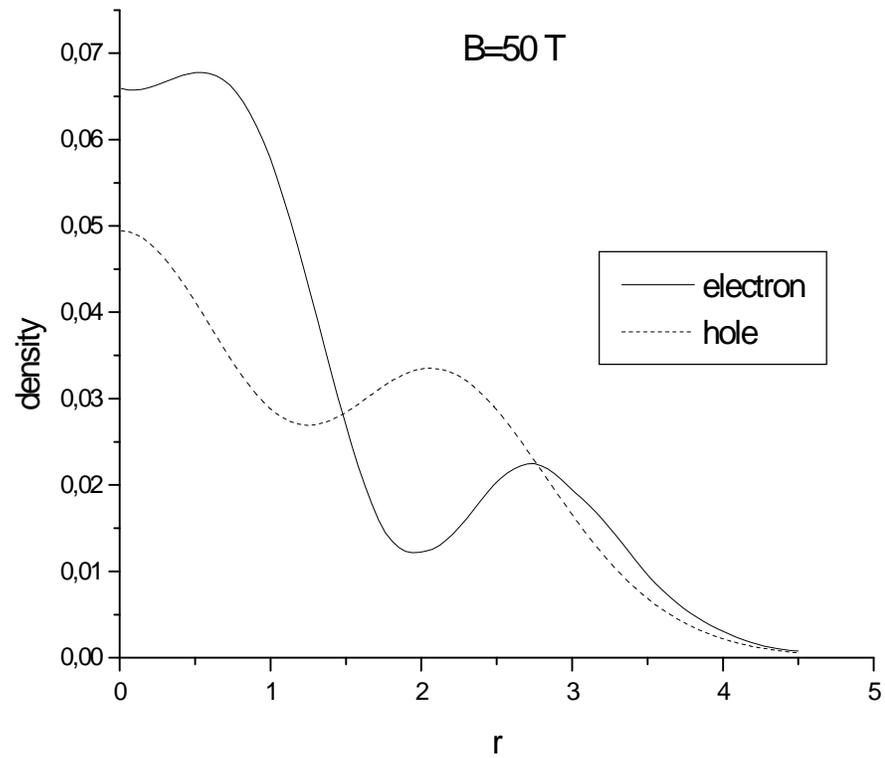
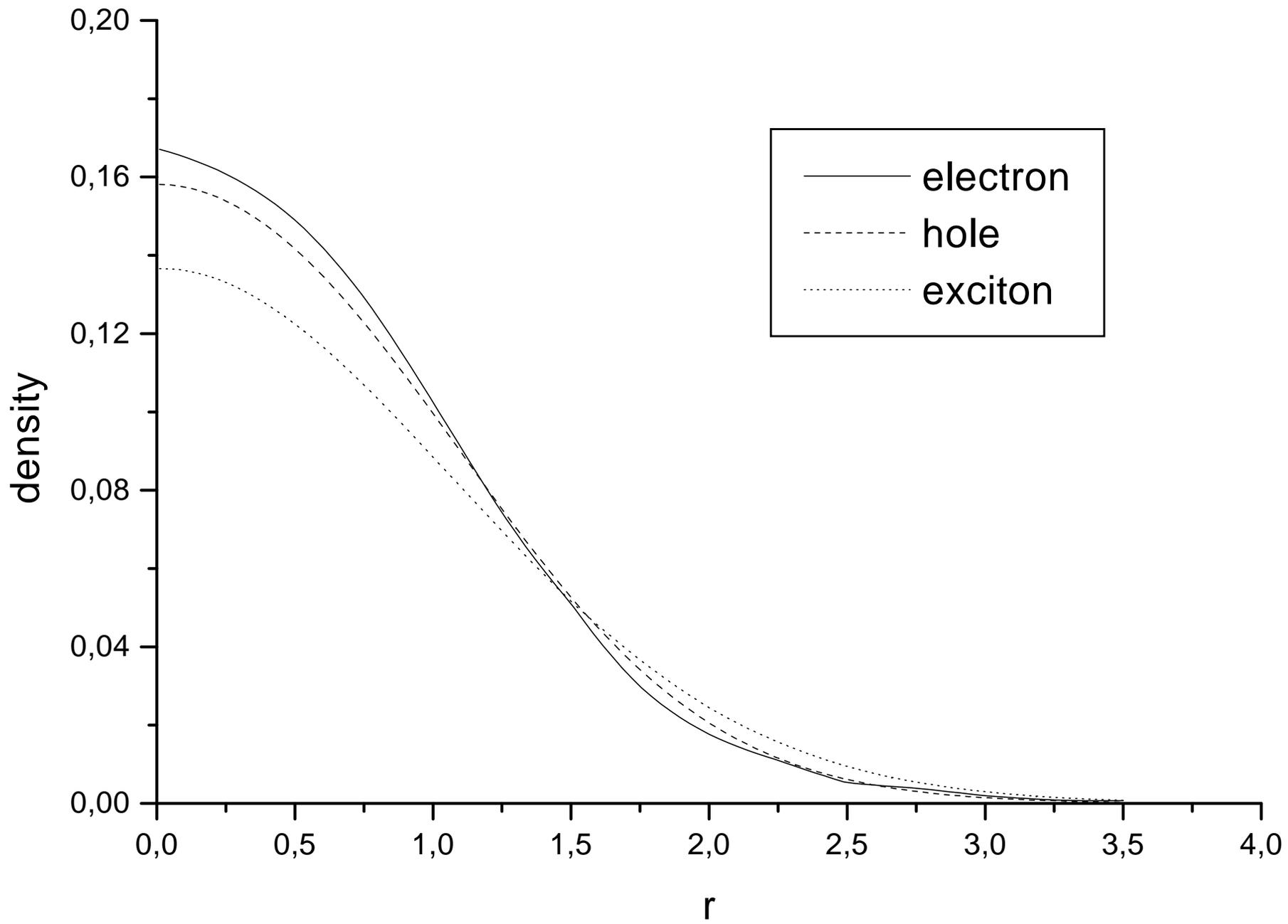
\end{document}